\documentclass[3p]{elsarticle}

\usepackage{amssymb}
\usepackage{amsmath}
\usepackage{lineno}
\journal{Applied Mathematics and Computation}

\begin{document}


\begin{frontmatter}

\title{Blocking defector invasion by focusing on the most successful partner}

\author[label1]{Attila Szolnoki}
\author[label2]{Xiaojie Chen}

\address[label1]{Institute of Technical Physics and Materials Science, Centre for Energy Research, Hungarian Academy of Sciences, P.O. Box 49, H-1525 Budapest, Hungary}

\address[label2]{School of Mathematical Sciences, University of Electronic Science and Technology of China, Chengdu 611731, China}

\begin{abstract}
According to the standard protocol of spatial public goods game, a cooperator player invests not only into his own game but also into the games organized by neighboring partners. In this work, we relax this assumption by allowing cooperators to decide which neighboring group to prefer instead of supporting them uniformly. In particular, we assume that they select their most successful neighbor and focus external investments exclusively into the related group. We show that this very simple alteration of the dynamical rule results in a surprisingly positive evolutionary outcome -- cooperators prevail even in harsh environment represented by small values of the synergy factor in the game. The microscopic mechanism behind the reported success of the cooperator strategy can be explained by a blocking mechanism which affects the propagations of competing strategies in a biased way. Our results, which remain intact by using different interaction topologies, reveal that it could be beneficial to concentrate individual efforts to reach a higher global wellbeing.
\end{abstract}


\end{frontmatter}

\section{Introduction}
\label{intro}

Explaining the emergence of cooperation among selfish agents who are interested in their best personal interest is a long-standing problem that has attracted intensive scientific activity in the last two decades \cite{szabo_pr07,perc_bs10,sigmund_10,szolnoki_jrsif14,cheng_f_pa19,murase_jtb18,takeshue_epl19,liu_dn_pa19}. Several mechanisms have already been identified, which highlight the importance of different forms of reciprocity \cite{nowak_s06}. As expected, monitoring players by rewarding positive act or punishing bad behavior supports cooperation efficiently \cite{sigmund_pnas01,szolnoki_epl10,cong_r_srep17,szolnoki_njp12,takesue_epl18}. Notably, the punishment can be executed in different ways like by decreasing payoff or via exclusion from mutual benefit \cite{szolnoki_pre17,liu_lj_srep17,quan_j_c19}.
In these cases the dilemma is transformed to another stage because not all cooperators want to bear the cost of an additional institution. Consequently, they become the so-called second-order free-riders and we practically face the original dilemma \cite{panchanathan_n04}. Interestingly, this problem can be resolved automatically in structured populations where players have limited range of interactions \cite{helbing_ploscb10,helbing_njp10}. In spatial systems, those players who bear the costs of both cooperation and additional institution can separate from simple cooperators hence the former group can fight against defection more efficiently \cite{szolnoki_pre11b}.

However, the most interesting intellectual challenge is to identify those rules and mechanisms which are strategy-neutral. In the latter cases, when we apply these rules, there is no obvious preliminary reason to support cooperative acts. An instructive example for a cooperator supporting environment is a highly heterogeneous interaction graph \cite{santos_prl05}. By following this research path, the breaking of symmetry, or the introduction of social diversity among competitors are now believed to be conducive to cooperation \cite{szolnoki_epl07,perc_pre08,rong_zh_c19}. For the mentioned cases, a generally valid explanation is that stronger players, who have higher social influence, can spread their strategies in their close neighborhood, resulting in a local coordination in the involved patches \cite{szolnoki_rsif15,yang_hx_csf17,szolnoki_njp18b,yang_hx_epl18}. Consequently, this type of separation of competing strategies reveals the evolutionary advantage of cooperation. 

Inhomogeneity can also be introduced in alternative ways. For example, in a public goods game, players are asked to contribute to a common pool. Their contributions are enlarged by a synergy factor and after it is redistributed among group members. In a structured population, a player is involved in different groups and cooperators are believed to contribute to all related games. However, there are real-life examples when players distribute their contributions in an unequal way. 

Although the option of heterogeneous investment into different games has been already suggested by some previous works, these research papers can be grouped along the following directions. In the first group, models focused on heterogeneous interaction graphs. Consequently, they assumed that a player's investment to an external group depends on the degree of the focal player \cite{cao_xb_pa10,wang_hc_pa18}. In this way, the suggested microscopic rule strongly utilizes the heterogeneous topology of interaction graph and does not consider the limited source of cooperator players. The other groups of models assume sophisticated and demanding skills of players. In particular, they suggest that a player's contribution to a specific group depends on their income received from the given group in the previous simulation step \cite{vukov_jtb11,zhang_hf_pa12}. Not really surprisingly, by applying such kind of microscopic rule, we practically enforce the reciprocity mechanism between cooperator players. It is because a cooperator group, where the redistribution is high, can expect additional support from an external cooperator. The rest of related models of heterogeneous investment broke the ``strategy-neutral" principle directly and considered actual strategy choice of group members. For example, in Ref.~\cite{yuan_wj_pone14} the contribution of a cooperator player into a group depends on the fraction of cooperators within that group. In an analogous work, it was suggested that the investment of a cooperator in a group organized by another partner depends on the reputation of the latter player \cite{yang_hx_pa19}. As expected, these rules resulted in a higher cooperation level because they directly support this strategy.

While our present model also considers heterogeneous investment in a public goods game, it does not belong to any of the previously mentioned paths. In particular, in our case, a player needs no additional information about the topology of interaction graph because his decision does not utilize the proper degree distribution of neighbors. Furthermore, players are not requested to record their previous incomes from specific groups for a decision about their investment. Instead they only need to know the payoff of their neighbors. The latter, however, should be available for every model, where strategy update is based on the payoff difference of interacting players. Last, and most importantly, our proposed microscopic rule is strategy-neutral, because when a player decides about his investment onto a specific group organized by one of his neighbors, then the actual strategy of his partner has no importance. Surprisingly, this very simple model, which requires nothing more than the traditional setup, provides a highly cooperative evolutionary income even at small values of synergy parameter which mimics harsh environment in general. 

In fact, we will show that without increasing the total investment of players it could be beneficial for the whole population to concentrate individual efforts instead of keeping the uniform, and seemingly more democratic investment policy. Before presenting our observations in more detail, we first proceed with the accurate description of the proposed public goods game with unequal investment rule.

\section{Focusing on the most successful partner}
\label{def}

For simplicity, we define the applied public goods game on a square lattice with periodic boundary conditions, but the extension to other topologies is straightforward. In the mentioned case, $L \times L$ players are arranged into overlapping groups of size $G = 5$ in a way that everyone is connected to its $G-1$ nearest neighbors. Consequently, each individual belongs to $g = 1, \dots G$ different groups where the first is organized by the focal player, while the rest $G-1$ games are organized by neighboring partners.

Initially, each player on site $x$ is designated either as a defector ($s_x = D$), or as a cooperator ($s_x = C$) randomly.
According to the standard protocol, a cooperator player invests a $c=1$ amount to each game while defectors contribute nothing. The sum of all contributions in each group is multiplied by the synergy factor $r$ and the resulting public goods are distributed equally amongst all group members independently of their strategies. Notably, the total payoff of every player is the sum of the incomes collected from related games. In the following we assume that a cooperator player considers an alternating investment policy with probability $\alpha$. We refer to this as selective cooperator ($SC$) state. Otherwise, with probability $1-\alpha$, he follows the standard investment protocol and invests into every external games equally. In the former case the cooperator invests into his own game the usual $c=1$ amount, but his remaining $(G-1) \cdot c$ contribution is distributed unequally. More precisely, the mentioned $SC$ player looks for the specific partner in his neighborhood with the highest payoff. After selective cooperator invests all his $(G-1) \cdot c$ external contribution exclusively to the game organized by the most successful neighbor. Given that there are more than one neighbors with the highest payoff value in the neighborhood then the mentioned cooperator selects one of them randomly.

In Fig.~\ref{def} we have summarized the investment policy of our present model. As mentioned, the focal {\it F} player collects income from not only his own game, which is marked by a yellow set, but also from the games organized by his neighbors. One of its external groups is marked by a dashed green ellipse in the plot. Evidently, the payoff collected from the game organized by the neighboring $n_1$ player depends on the contribution of $m_1$ player. If this player is in a $SC$ state, then $m_1$ contributes to $n_1$'s game only if the payoff of $n_1$ player exceeds the payoff values of $k_1, k_2$, and $k_3$ neighboring players. In the mentioned case $m_1$ contributes to $n_1$'s game by $(G-1)\cdot c=4$ amount, as illustrated by an arrow. Otherwise it contributes nothing, no matter $m_1$ is a cooperator. Notably, a cooperator player always contributes a $c=1$ amount into his own game independently whether he is in an unconditional or selective cooperator state.

\begin{figure}
\centering
\includegraphics[width=6.5cm]{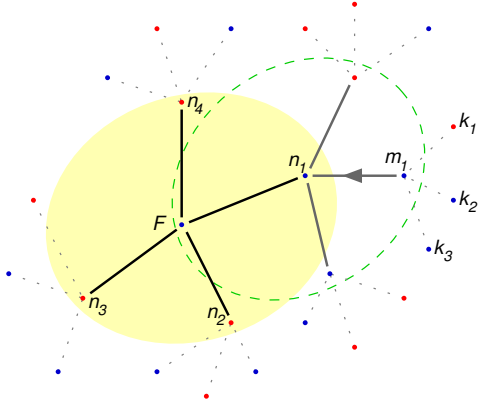}\\
\caption{The focal {\it F} player collects income from not only his own game, marked by a yellow set, but also from the game organized by his neighbor $n_1$. The group of the latter game is marked by a dashed green ellipse. Note, however, if cooperator $m_1$ is focusing on the best neighbor, then he invests contribution to $n_1$'s game only if the payoff of $n_1$ player is higher than the payoff of $k_1, \dots, k_3$ players. If this is the case then $m_1$ player invests all his external $(G-1) \cdot c=4$ contribution here. The latter act is marked by an arrow. Similarly, the focal {\it F} cooperator always contributes his own game, but his external investment depends on its own state. In normal case, that happens with probability $1-\alpha$, a regular cooperator $F$ contributes to $n_1$'s game by $c=1$. With probability $\alpha$, $F$ focuses on the best neighbor and invests into $n_1$'s game only if the payoff of $n_1$ exceeds the payoff of $n_2, \dots, n_4$ players. Otherwise, $F$ contributes nothing to $n_1$'s game no matter he is in a cooperator state.}\label{def}
\end{figure}

We should stress that contrary to previous works, our modified investment rule does not utilize the heterogeneous topology of interaction graph, hence it can be applied for homogeneous topology as well. We also note that our observations are not restricted to lattice topology, but remain valid on random graphs too. We note that since the investment decision of a selective cooperator requires the knowledge of actual payoff values of neighbors, therefore in the zero step we provide a random payoff value for all players from the $[G \cdot (r-1) / 2 \pm 1]$ interval. Naturally, in the following steps the players' payoff values are updated according to the proper states of their neighbors. We highlight that the actual initial payoff values have no relevant consequence on the outcome, they only serve the proper launch of the simulation steps. When we modified these values, we observed identical final state at a specific values of $r$ and $\alpha$.

The rest of the dynamical rule follows the standard procedure. More precisely, during an elementary Monte Carlo step we choose a player $x$ and one of his nearest neighbors $y$ at random. If the strategies of these players are different, then the related $\Pi_x$ and $\Pi_y$ payoff values are calculated by summing all the incomes acquired in each individual group. Then player $y$ adopts the strategy from player $x$ with a probability given by the Fermi function
$w=\{1+\exp[(\Pi_y-\Pi_x)/K]\}^{-1}$, where $K=0.5$ quantifies the uncertainty by strategy adoptions \cite{szabo_pre98}. A full Monte Carlo step offers a chance to every player to change his strategy once on average.

The system size for the square grid was varied from $L \times L = 100 \times 100$ to $400 \times 400$, and we have not observed relevant finite-size effect. As we will stress, our model is capable to support cooperation even in homogeneous graphs where players have identical degree. This is a striking difference from those previous cases where the proposed asymmetry applied in the microscopic rule utilized the inhomogeneity of interaction graphs intensively \cite{cao_xb_pa10,zhang_hf_pa12,yuan_wj_pone14,wang_hc_pa18}. To broaden the robustness of our observation, we will leave square-lattice topology and also consider a random graph where there is no translation invariance, but small-world character emerges. This can be done by using a regular random graph where links of an initial lattice are rewired hence the degree distribution remains uniform \cite{szabo_jpa04}. In the latter case, when random interaction graph was applied, we monitored typically $N=10^5$ players. In all cases, the stationary value of cooperation level $f_C$ was determined after a typical $10^4$ relaxation steps and $f_C$ values were averaged 50 independent runs. In the light of results presented first in the next section, we have also studied a slightly modified model which details will be described in the next section.

\section{Results}

\begin{figure}[h!]
\centering
\includegraphics[width=7.5cm]{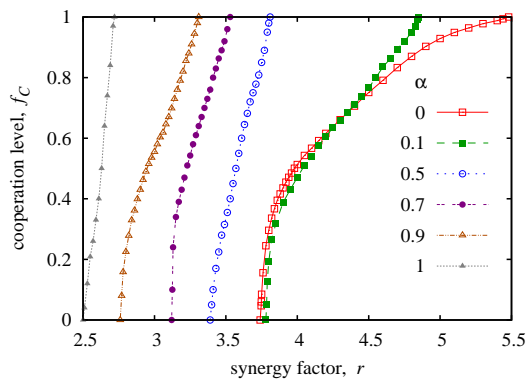}\\
\caption{Cooperation level in dependence on synergy factor for different values of $\alpha$ parameter as indicated in the figure legend. This plot suggests that cooperation is largely supported if cooperator players prefer to support exclusively the best neighbor independently of the latter strategy. The results were obtained on square lattice where the linear system size are $L=300$ and the error bars are comparable to the symbol size.}\label{sqr_r}
\end{figure}

We first present some representative cooperation levels obtained at different $\alpha$ values in dependence of the synergy factor. As Fig.~\ref{sqr_r} highlights, the introduction of unequal investment policy stimulates a positive evolutionary outcome. More precisely, by increasing $\alpha$, means when we increase the chance that a cooperator behaves as a selective cooperator, then cooperators survive even at small values of synergy factor $r$. For example, at $r\approx 3.7$ and $\alpha=0.5$, when there is equal chance that a cooperator distributes his contributions uniformly or focuses them into a single group, then the system evolves toward a full cooperator state. Note that at this value of synergy factor the system would always terminate into a full defector state at $\alpha=0$, when cooperators disseminate their contributions among their neighbors uniformly.

\begin{figure}
\centering
\includegraphics[width=7.5cm]{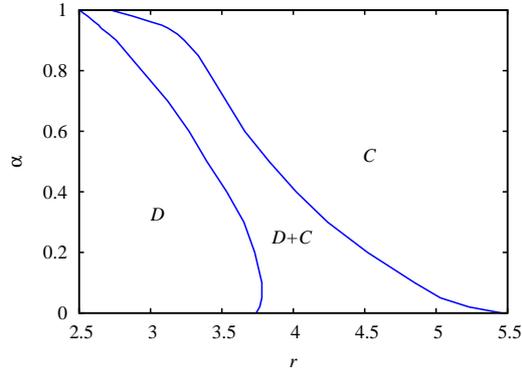}\\
\caption{Phase diagram, depicting the stable solutions ($C$-full cooperator state, $D$-full defector state and $D+C$-mixed state) on the $\alpha-r$ parameter plane. In agreement with the previous plot, cooperators can fight more efficiently against defection at high $\alpha$ values, where they focus their external contributions on a single neighbor, who does the best in their neighborhood.}\label{phd}
\end{figure}

The complete behavior on the $\alpha-r$ parameter plane is summarized in Fig.~\ref{phd}. This plot confirms that a higher cooperation level can be achieved when cooperator players prefer to focus their external investment toward a specific neighbor, instead of supporting the whole environment uniformly. The best results are obtained when cooperators give up traditional unconditional cooperator state and instead they select their investment target exclusively. We stress that we do not expect any additional effort or cognitive skill from players, which are not already available in the traditional model. In particular, players do not need anything to know about the interaction topology, which was a fundamental condition in previous works. Furthermore, they do not have to record their past income originated from earlier games played with their neighbors. Evidently, to collect the mentioned information would require an extra effort that should be considered via an extra cost or lowered payoff value. But in our present model, when cooperators make a decision about their investment, they only need to know the payoff of their neighbors. This information, however, is available in the traditional model because imitation probability is also based on the payoff difference of competing partners.

We stress that our observations are not limited to square lattice but remain valid for other topologies. Evidently, if the interaction graph is highly heterogeneous, then hubs, who are able to collect high payoff, will successfully attract the investments of neighboring cooperators. Consequently, in this case we can observe the same mechanism as was previously observed for scale-free graphs \cite{santos_n08}. Namely, cooperator hubs will collect high payoff and become strong. Initially, defector hubs can utilize their neighbors, but later, when the neighboring players adopts the most successful strategy, then defector hubs become vulnerable. In this way it is not really surprising that heterogeneous investment supports cooperation on highly heterogeneous graphs \cite{cao_xb_pa10,wang_hc_pa18}.

\begin{figure}[h!]
\centering
\includegraphics[width=7.5cm]{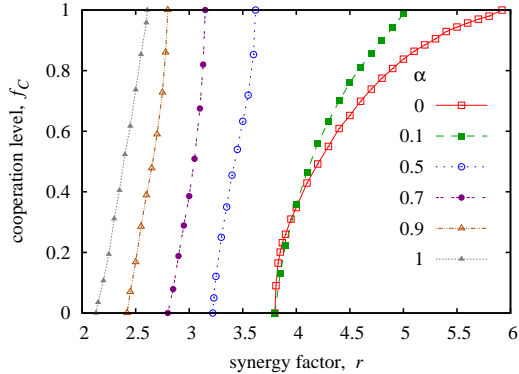}\\
\caption{Cooperation level in dependence on synergy factor for different values of $\alpha$ parameter as indicated in the figure legend. Here random interaction graph was applied where we used the same $k=4$ degree as for square lattice. The cooperator supporting consequence of the selected investment protocol remained intact no matter random topology was used. The system contained $N=10^5$ players.}\label{rrg}
\end{figure}

From this point of view, it is more interesting to check homogeneous graphs where there is no relevant difference between the degree distribution of players, but the topology is not necessarily translation invariant as for square lattice. Motivated by these arguments, in Fig.~\ref{rrg} we present results obtained by using random regular graph. For proper comparison  we used the same $k=4$ degree distribution as for square grid \cite{szabo_jpa04}. This plot suggests very similar behavior we previously reported in Fig.~\ref{sqr_r}. Consequently, the positive impact of selected investment on cooperation level is a more general phenomenon that is not restricted to translational invariance interaction topologies. Hence we can conclude that the reported effect can be observed even on homogeneous graphs where there is no significant difference in players' degree. But more importantly, we do not need a highly heterogeneous degree distribution which was an essential condition to obtain a cooperation supporting mechanism previously.

Until this point we assumed that cooperators were uniform and they all have a certain chance to switch from traditional cooperator into the selective cooperator state, which is controlled by parameter $\alpha$. The phase diagram, however, illustrates that the best solution can be reached at $\alpha=1$, which means that cooperators always behave as selective cooperator in the latter case. An intriguing question can be posed, here. Namely, is it possible to reach a higher global well-being via a selection mechanism? Put differently, how does the system evolve if we introduce pure cooperators and selective cooperators as permanent strategies simultaneously? In the latter case we have a three-strategy model where unconditional defection ($D$), unconditional cooperation ($C$) and selective cooperation ($SC$) strategies compete. Their relation is far from trivial because a $C$ and a $SC$ player have identical cost, the only difference is how they distribute their investment. Furthermore, an $SC$ supports just only one of his neighbors, therefore the emergence of networks reciprocity, which is a basic mechanism in structured populations, can hardly evolve among $SC$ players. To clarify this question we have studied this modified model and found that unconditional cooperators always die out, hence the evolutionary outcome depends only on the relation of $D$ and $SC$ strategies. In other words, the system always terminates into the state that was observed for the $\alpha=1$ case in the previously studied homogeneous model. Therefore we can conclude that $SC$ strategy, which is globally beneficial, could be selected during an evolutionary process.

\begin{figure}[h!]
\centering
\includegraphics[width=3.95cm]{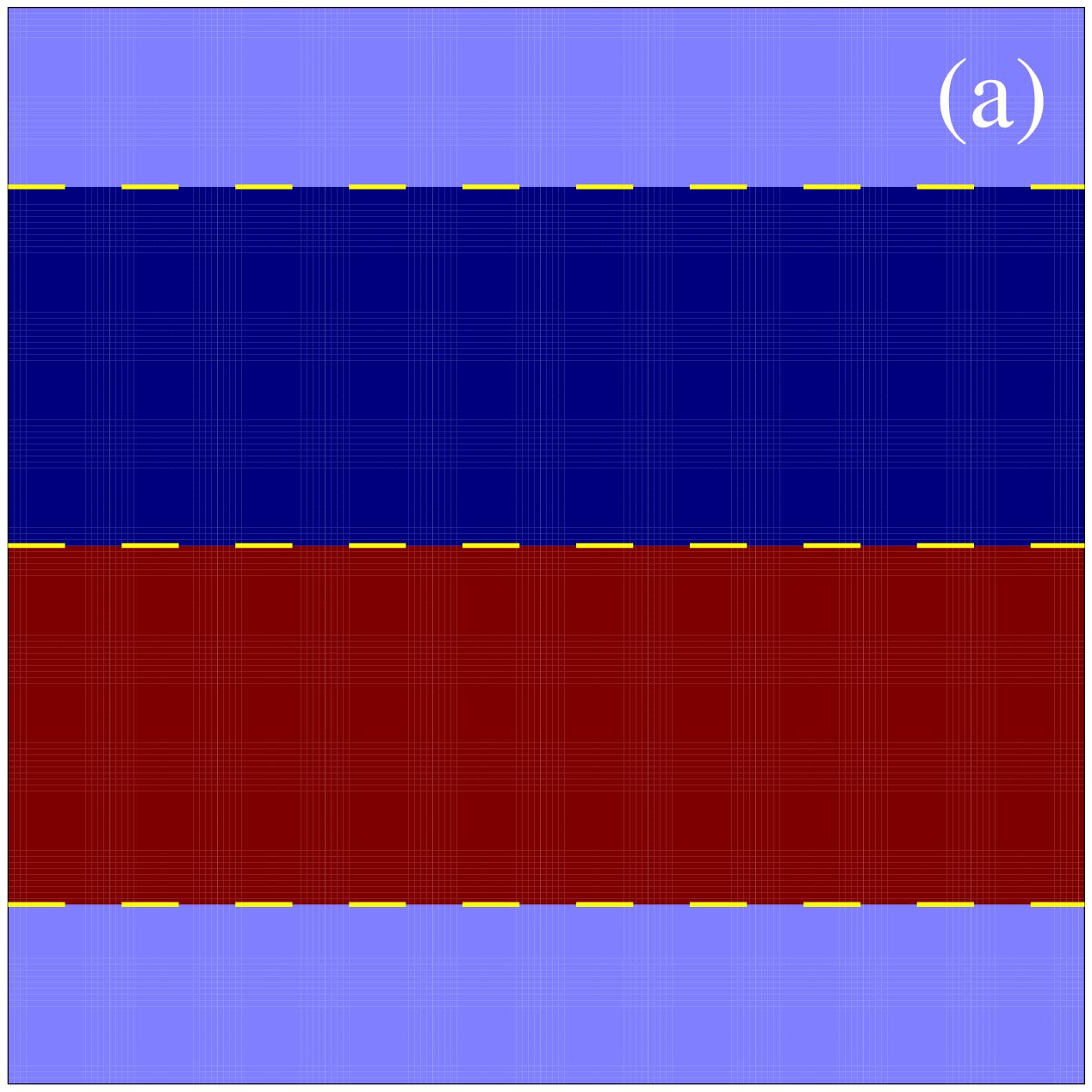}\includegraphics[width=3.95cm]{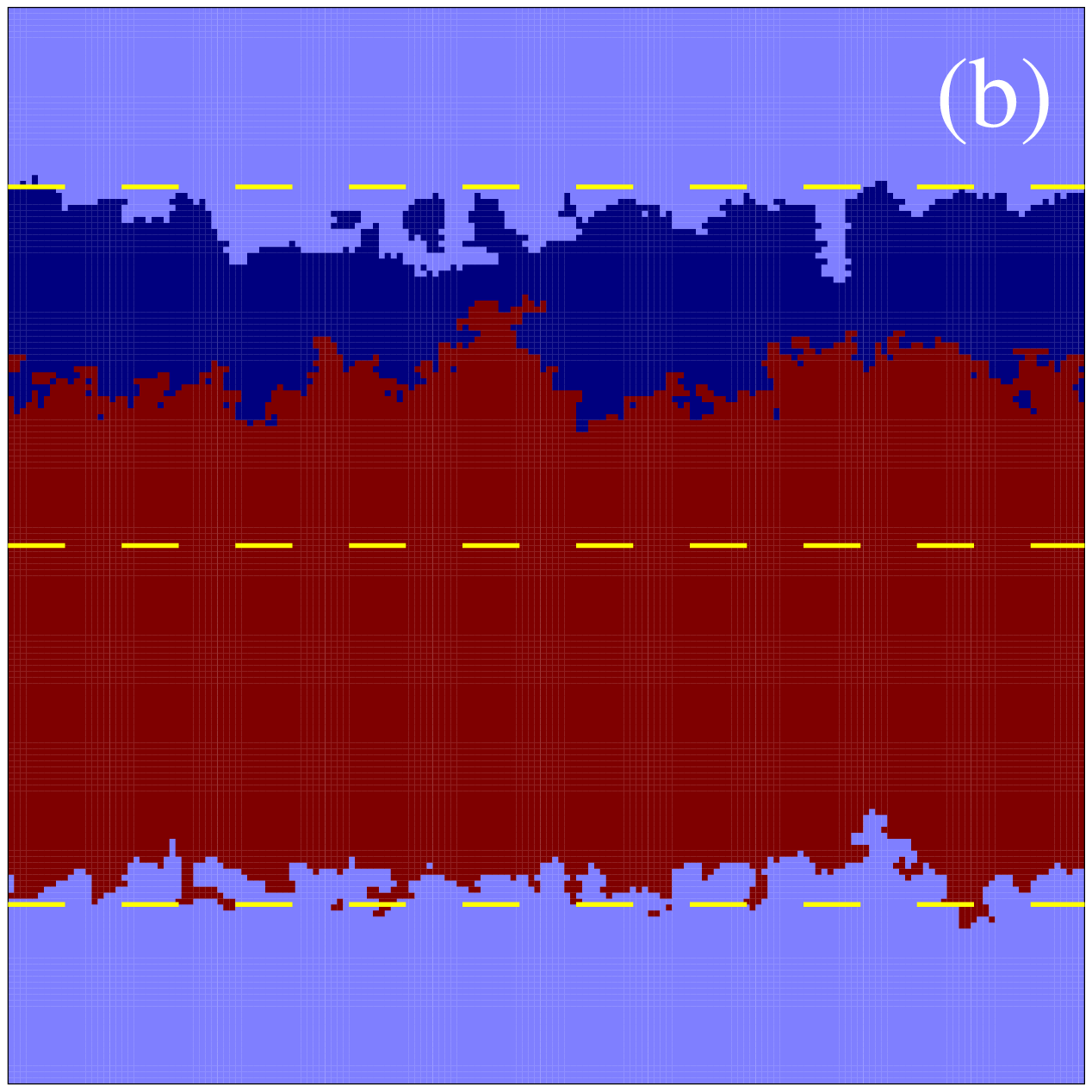}\includegraphics[width=3.95cm]{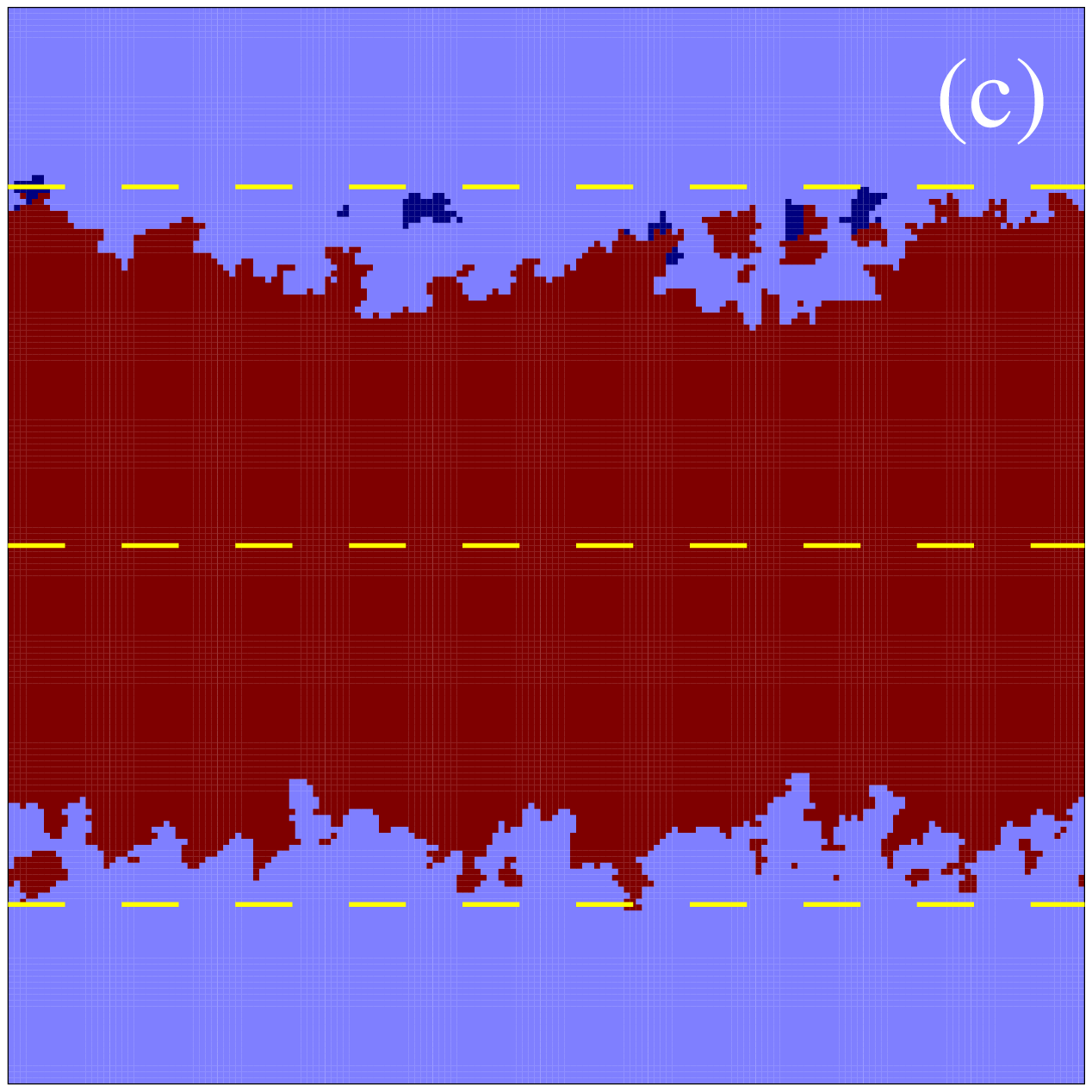}\includegraphics[width=3.95cm]{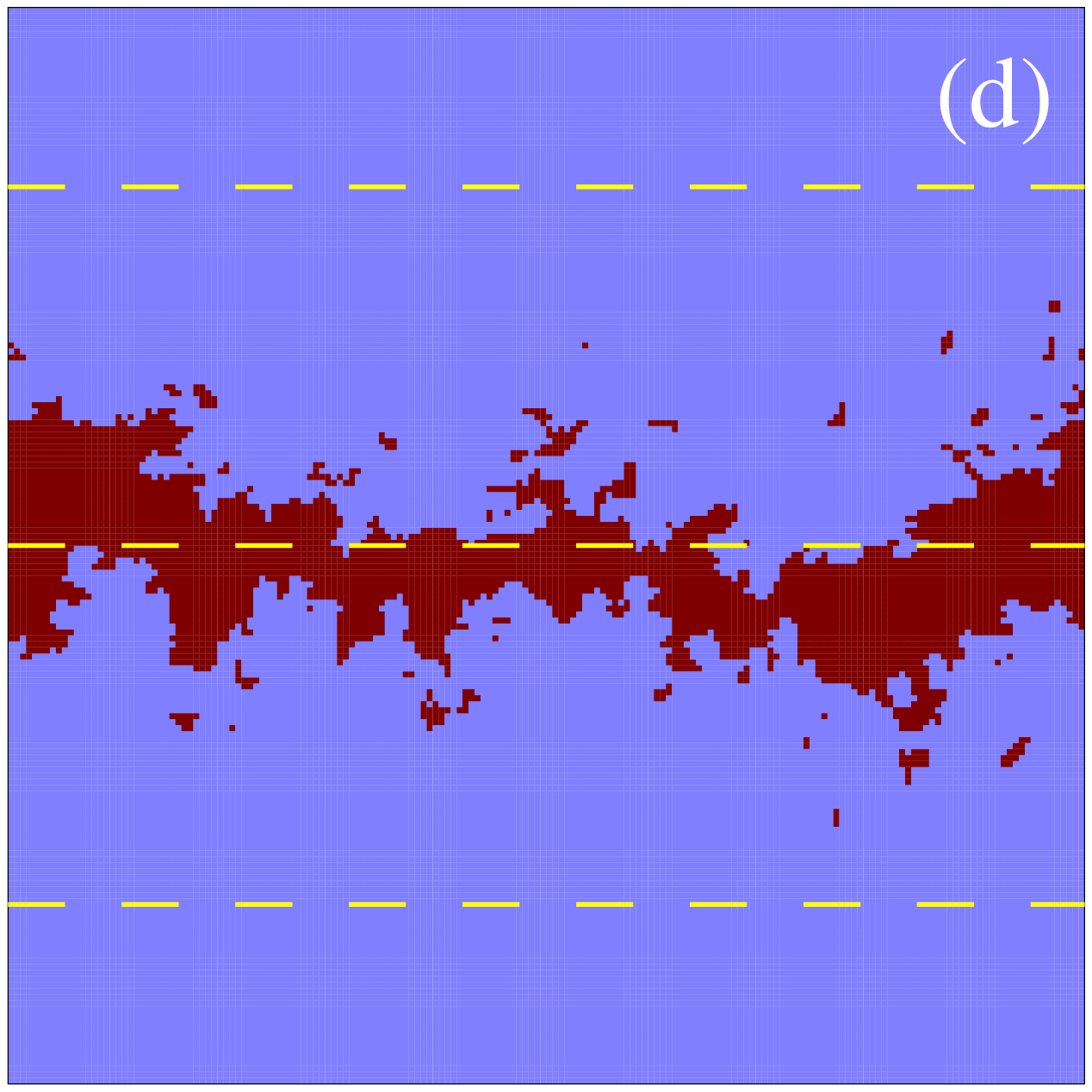}\\
\caption{Characteristic snapshots of pattern formation starting from a prepared initial state. Here we have three pure strategies, namely defectors (red), unconditional cooperators (dark blue), and selective cooperators (light blue). The initial positions of frontiers are marked by dashed yellow lines in all panels. Panel~(b) illustrates clearly that both defectors and selective cooperators invade unconditional cooperators. But defectors, shown in panel~(c), do it more efficiently and conquer the majority of space originally occupied by unconditional cooperators. Selective cooperators, however, dominate defectors and finally prevail the whole system (not shown). The snapshots of $180 \times 180$ system were taken at 0 (a), 100 (b), 200 (c) and 700 (d) full $MCS$ steps, and the synergy parameter value was $r = 3.0$.}\label{L180}
\end{figure}

To understand the mechanism more deeply, which is responsible for the advantage of $SC$ strategy, in Fig.~\ref{L180} we present a pattern formation process when the system was launched from a prepared initial state. Here the three competing strategies are designated by different colors. Namely, defectors are marked by red, unconditional cooperators by deep blue while selective cooperators are denoted by light blue. Intentionally, we here choose a low $r=3$ value of synergy factor that would result in a full defection in the traditional model. For easier comparison we marked by dashed yellow lines the original positions of border lines. In panel~(b) and in panel~(c) it is easy to see that both defectors and $SC$ players invade unconditional cooperators. But the invasion of $D$ strategy is more effective and they invade the majority of space originally occupied by $C$ strategy. When $C$ players die out, the proper final state of the evolution depends only on the relation of $D$ and $SC$ strategies. At the mentioned specific $r=3$ value of synergy factor, as already shown in Fig.~\ref{phd}, $SC$ invade defectors and prevail the whole system. Note that this final state is not shown in the plot.

\begin{figure}
\centering
\includegraphics[width=3.95cm]{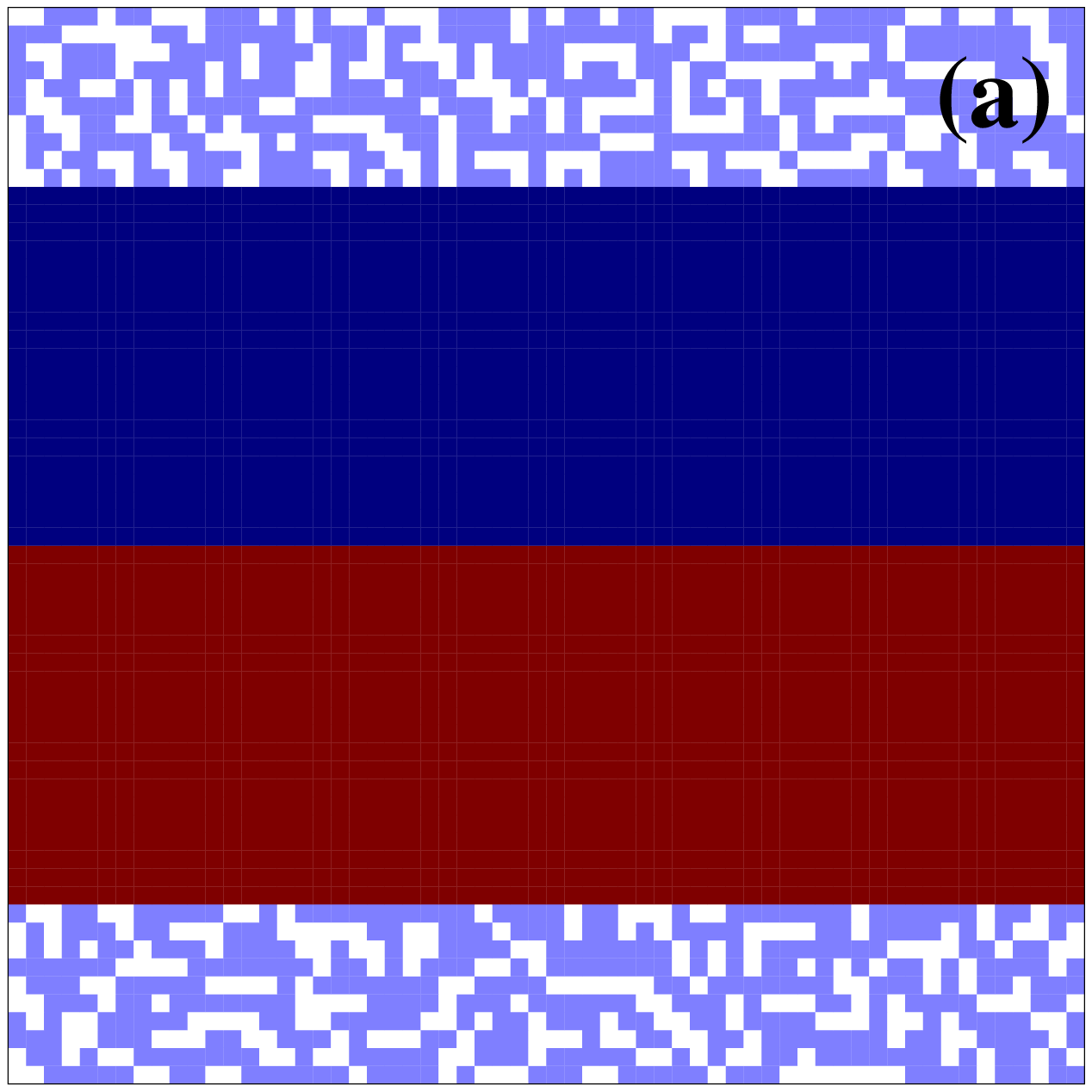}\includegraphics[width=3.95cm]{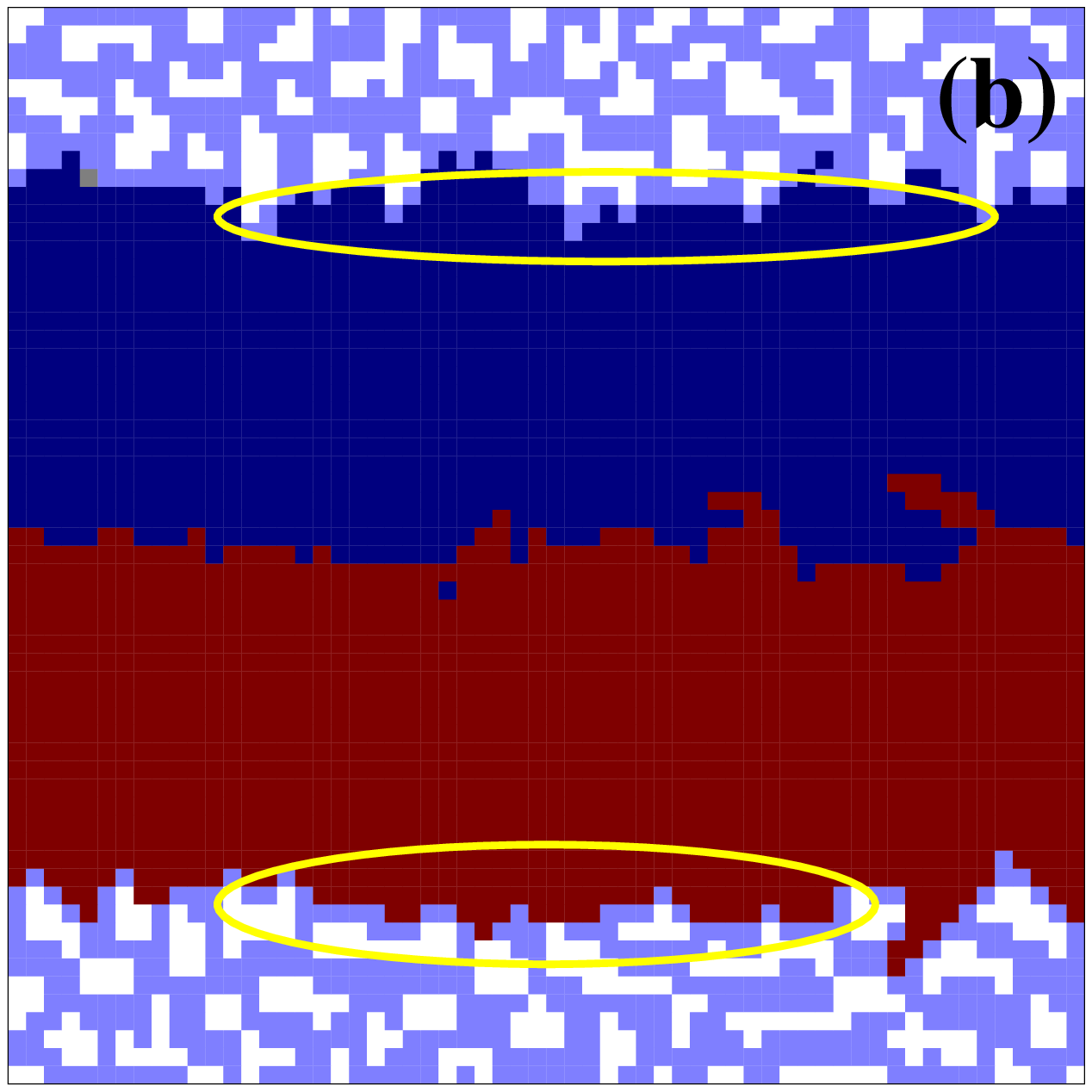}\includegraphics[width=3.95cm]{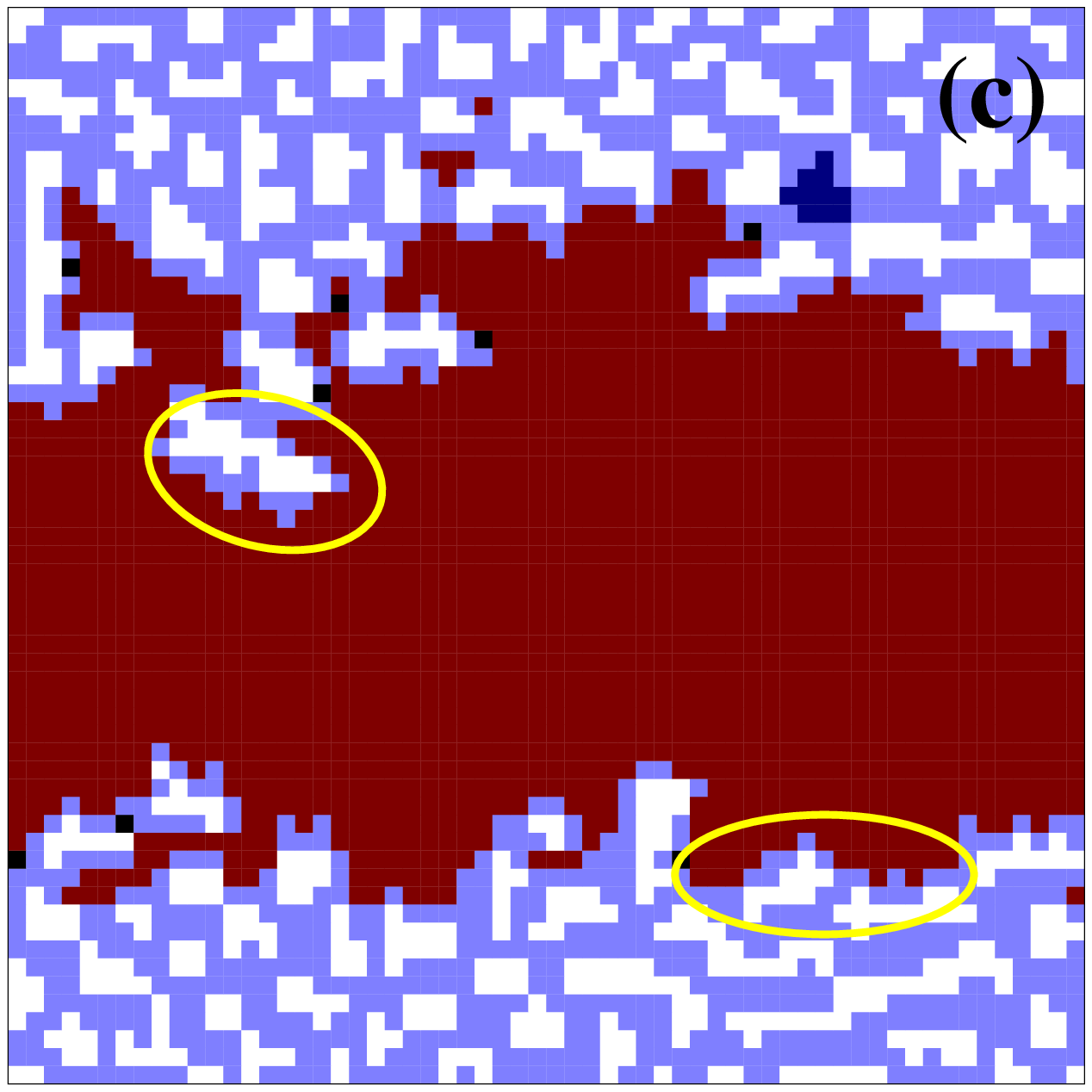}\includegraphics[width=3.95cm]{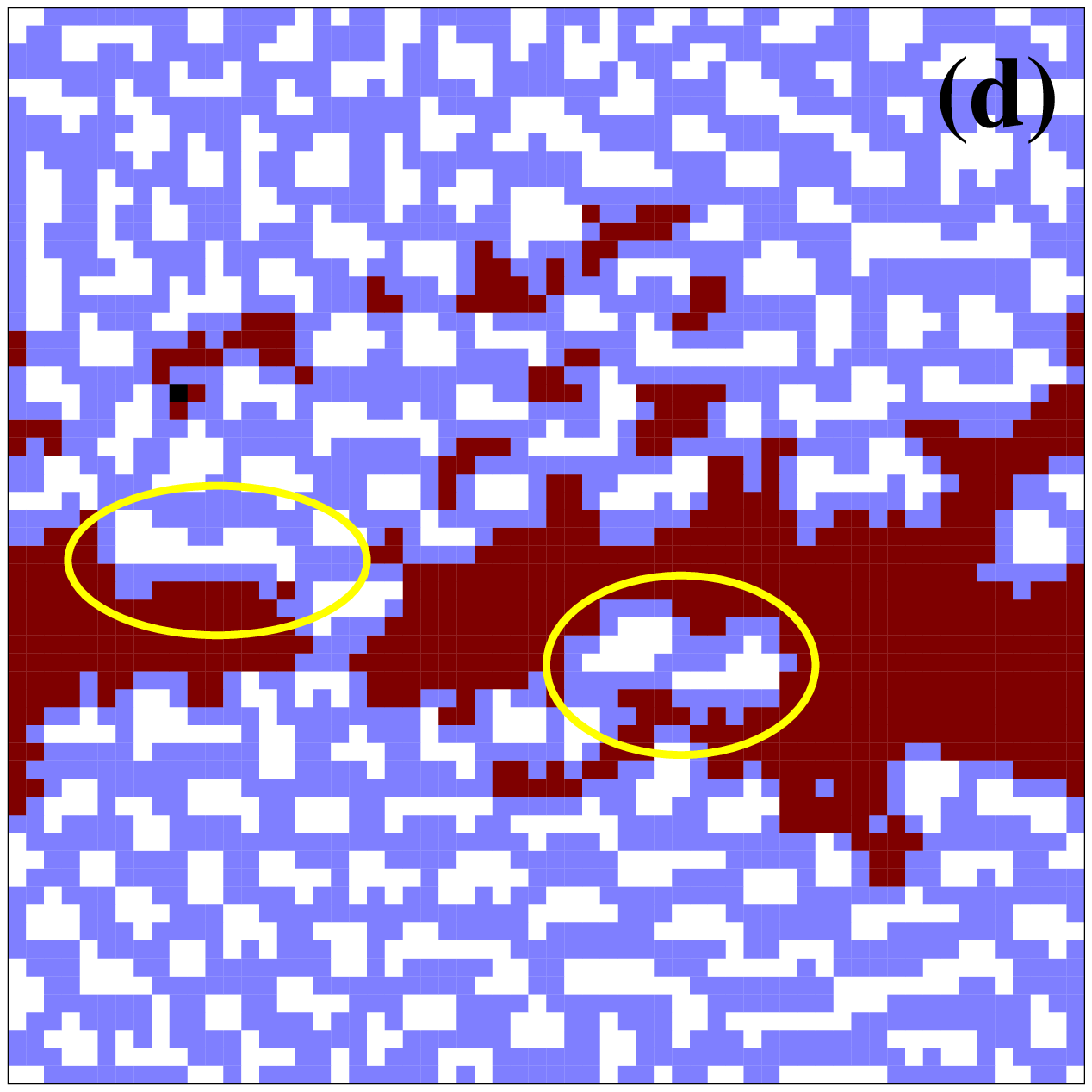}\\
\caption{Pattern formation starting from the same initial setup as for Fig.~\ref{L180}. Here we used a special coloring to mark those players who have the highest payoff in a group and enjoy the support of a neighboring selective cooperator player. More specifically, we mark those defectors, unconditional cooperators, and selective cooperators by black, gray, and white color respectively. For clarity we present a smaller $60 \times 60$ system where individual players are visible. The mentioned supported players are not present in the borderline between competing domains, but they are generally behind it in the next lines. Their typical positions during the evolution are highlighted by yellow ellipses in panels~(b-d).}\label{L60}
\end{figure}

For a full explanation, it is instructive to apply an alternative coloring technique that reveals the representative microscopic process. For this purpose, we use not just the previously introduced colors of strategies but we also add three extra colors. Namely, we mark by black, gray, and white colors those $D, C$, and $SC$ players respectively who are supported by a neighboring $SC$ player with an extra large $(G-1)\cdot c$ investment. This coloring technique is applied in Fig.~\ref{L60} where we present a smaller system to make individual players visible. Similarly to the previous plot here we used the same $r=3$ parameter value, therefore the system will terminate again into the full $SC$ state (not shown in the figure). In panel~(a) we only see white ``highly supported" players who are distributed randomly in the bulk of $SC$ domain. This is due to the originally randomly distributed payoff values in the zero step. Later, when the strategy propagation is launched, fronts start moving between homogeneous domains. Interestingly, however, it is very rare that black or gray pixels emerge, which simply means that it almost never happens that an $SC$ player supports a neighboring $C$ or $D$ player. Furthermore, which is also very important, white pixels cannot be detected in the front line. More precisely, at the front separating $D$ and $SC$ domains, it is typical that red and light blue players are facing each other. Similar situation can be detected at the front between $C$ and $SC$ domains, where black and white pixels are hardly detected. The representative feature of the emerging patterns are highlighted by yellow ellipses in the panels~(b--c) of Fig.~\ref{L60}.

\begin{figure}[h!]
\centering
\includegraphics[width=2.5cm]{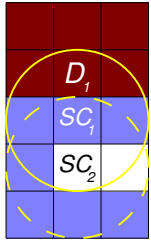}\\
\caption{Mechanism that blocks the invasion of defector strategy. Here we used the same color coding as for previous Fig.~\ref{L60}. Accordingly, defector and selective cooperator domains compete where white color marks those players who are supported exclusively by neighboring $SC$ partners. $D_1$ cooperator at the border cannot utilize the selective cooperator $SC_1$ player because $SC_1$ donates all his external contribution to the game organized by $SC_2$ player. Furthermore, $D_1$ has just a minimal benefit from the game organized by $SC_1$ player because here the only contribution comes from $SC_1$. This group is marked by a solid yellow curve. On the other hand, $SC_1$ is able to collect a reasonable income from the game organized by the $SC_2$ player. The latter group is marked by a dashed yellow circle. As a result, $SC_1$'s cumulative payoff can be larger than the payoff of $D_1$ player even for relatively small $r$ values.}\label{block}
\end{figure}

Based on these observations we can easily reveal the key mechanism that is responsible for the success of $SC$ strategy. For simplicity, we describe the elementary step of domain wall propagation between $D$ and $SC$ domains, but conceptually similar explanation can be given for the competition of $C$ and $SC$ strategies. In Fig.~\ref{block} we show the competing domains, where neighboring $D_1$ and $SC_1$ compare their payoff during the strategy invasion. In the traditional model $D_1$ would exploit a neighboring cooperator partner. First, because an unconditional cooperator would invest directly to the game organized by $D_1$. Secondly, $D_1$ would largely benefit from the game organized by the mentioned cooperator because a substantial contribution is collected from cooperator members of the mentioned group. But now both elements are missing. First, $SC_1$ does not invest directly into the game organized by $D_1$ because $SC_2$ is more attractive target by having a higher payoff. Secondly, the income from the game organized by $SC_1$ is minimal. It is because only $SC_1$ invests into his own game marked by a solid yellow circle, all other neighboring selective cooperator support a more successful neighbor. In sum, $D_1$ has only a modest income in the neighborhood of an $SC$ player.

As we argued, $SC_1$ is weakened at the front, but he has an escape route not to be totally vulnerable. It is because $SC_1$ benefits from the very successful game organized by the neighboring $SC_2$ player. This game is marked by a dashed yellow circle in the plot. Here not just $SC_1$, but also other neighboring $SC$ players invest a huge amount into the game, hence all enjoy a large split after multiplication. As a result, $SC_1$ still has a reasonable payoff that is competitive to the payoff of $D_1$. Summing up, the introduction of a selective investment policy results in weakened players in both sided of the front, but defectors suffer more, hence their invasions are largely blocked.

It is easy to see that similar argument can be given for the competition of $C$ and $SC$ strategies, which explains why $SC$ invades unconditional cooperation no matter they bear the same total cost and the emergence of network reciprocity is less obvious for the former strategy.

\section{Conclusion}

It is clear to see that when someone cooperates then the primary aim is to elevate the general well-being of the group or community. But one may ask which way serves this goal more efficiently? Is it better to support anyone or is there a smarter way to use our efforts? For instance, one may argue that it is useless to support a group which does not functioning well, where the leader is unsuccessful. On the other hand, our investment may reach the highest impact if we focus on a venture that is organized by a successful player. In this work we have elaborated this idea where we allowed cooperator players to concentrate their external investments to a single group led by the most successful neighbor. Importantly, a cooperator does not consider the strategy of the target neighbor, but checks only its payoff. In this way the decision about a cooperator's investment requires no extra information comparing to the standard public goods game. Surprisingly, the suggested strategy-neutral investment policy elevates the cooperation level dramatically even at small values of synergy factor, where the traditional model would suggest a full defection state.

We note that different forms of heterogeneous investment was already studied by several earlier works \cite{cao_xb_pa10,wang_hc_pa18,fan_rg_pa17}. However, they assumed more complicated rules which demand more intellectual effort, hence additional care from cooperator players. Our present model, however, is the simplest because it assumes nothing additional information than is already available for players during the strategy imitation process. It is worth stressing that the present positive effect also works in homogeneous graphs where some earlier models, which built on the strong heterogeneity of the interaction graph, would fail to support cooperation.

The microscopic mechanism, which explains the success of the suggested protocol, is based not on the usual reciprocity-based arguments. The latter can be found in various forms in models of structured populations, where limited number of interactions of players offers not just the chance to separate from those who exploit others but also enlarge the positive consequence of direct reciprocity \cite{allen_pre18,shao_yx_epl19,fotouhi_rsif19,wang_x_rspa20}. In our present case, as we argued, the introduced investment policy weaken all fighters who are in the front line between the competing domains - independently of their actual strategies. But this weakening effect is biased and defectors suffer more from it. As a consequence, they are unable to exploit the vicinity of cooperators hence they loose their evolutionary advantage. They become less attractive and their invasion is completely blocked. The mentioned weakened cooperators in the front line, however, still have a chance to enjoy the vicinity of successful cooperators behind them, hence they benefit modestly from the success of their neighbors. In sum, weakened cooperators still do better than weakened defectors, hence the direction of strategy propagation can be reversed.

It is worth noting that the reported cooperator supporting mechanism fits nicely to those observations where the introduced strategy-neutral rule has biased impact on the strategy invasion of competing strategies \cite{szolnoki_pre09,du_wb_ctp12,szolnoki_pre18,chang_sh_pa18,szolnoki_srep19,fu_mj_pa19}. Hence these mechanisms provide an alternative way to understand to original enigma and explain why cooperation may prevail among selfish agents.

\vspace{0.5cm}

This research was supported by the Hungarian National Research Fund (Grant K-120785) and by the National Natural Science Foundation of China (Grants No. 61976048 and No. 61503062).

\bibliographystyle{elsarticle-num-names}

\end{document}